\documentclass[aip,jcp,reprint]{revtex4-1}
\usepackage{epsfig}
\usepackage{color}
\usepackage{amsmath}
\usepackage{amsfonts}
\usepackage{natbib}
\usepackage{notes2bib}

\newcommand{\degree}{^{\circ} }

\newcommand{\avg}[1]{\left<#1\right>}

\newcommand{\brac}[1]{\left[#1\right]}

\newcommand{\kT}{\ensuremath{k_{\rm B}T}}

\begin{document}


\title{Molecular Simulation of Covalent Bond Dynamics in Liquid Silicon}


\author{Richard C. Remsing}
\email[]{rick.remsing@rutgers.edu}
\affiliation{Department of Chemistry and Chemical Biology, Rutgers University, Piscataway, NJ 08854}
\author{Michael L. Klein}
\email[]{mike.klein@temple.edu}
\affiliation{Institute for Computational Molecular Science and Department of Chemistry, Temple University, Philadelphia, PA 19122}




\begin{abstract}
Many atomic liquids can form transient covalent bonds reminiscent of those in the corresponding solid states.
These directional interactions dictate many important properties of the liquid state,
necessitating a quantitative, atomic-scale understanding of bonding in these complex systems.
A prototypical example is liquid silicon, wherein transient covalent bonds give rise to local
tetrahedral order and consequent non-trivial effects on liquid state thermodynamics and dynamics.
To further understand covalent bonding in liquid silicon, and similar liquids,
we present an ab initio simulation-based approach for quantifying
the structure and dynamics of covalent bonds in condensed phases.
Through the examination of structural correlations among silicon nuclei and maximally localized
Wannier function centers, we develop a geometric criterion for covalent bonds in liquid Si. 
We use this to monitor the dynamics of transient covalent bonding in the liquid state and estimate
a covalent bond lifetime. 
We compare covalent bond dynamics to other processes in liquid Si
and similar liquids and suggest experiments to measure the covalent bond lifetime. 
\end{abstract}


\maketitle

\raggedbottom

\section{Introduction}

Many metallic and semi-metallic atomic liquids contain significant numbers of dynamic covalent bonds
reminiscent of the static bonds formed in the corresponding solid state.
The covalent bonds in these liquids are dynamic, readily breaking and forming
on molecular timescales, and the characterization of these processes is complicated 
due to the interplay between electronic and nuclear structures.
Such metallic liquids include molten silicon~\cite{Stich:PRB:1991,Okada:2012fb,Stich:PRL:1996,Stich:PRL:1989,Ashwin:2004oq,SCANNature,SiPRB,Remsing_2018}, boron~\cite{Okada_2015},
gallium~\cite{Gong_EL_1993}, and hydrogen at high pressure and temperature~\cite{Rillo_2019,Morales_2013,Pierleoni_2017,Pierleoni_2016,Pierleoni_2018,Zaghoo_2017},
as well as many alloys, including those
of importance in phase change random access memory materials~\cite{lee2017the,Loke1566,Skelton_2015}.
These liquids play important roles in fuel cells, catalysis, and electrochemistry, where the dynamic
covalent bonds are expected to play an importance role in chemical reactivity~\cite{Zavabeti332}.
Many of these liquids are also found in planetary cores~\cite{Poirier:1994,Pozzo_2013,Pozzo:2014,Zaghoo_2017,Williams:2018,Mazzola_2018}
and understanding their structure and dynamics is of importance to planetary and geophysical sciences.
In all of these fluids, a complete understanding of their properties requires knowledge of the fundamental
interactions and timescales governing their chemical and physical transformations. 
The relative abundance of dynamic covalent bonds is expected to play a role in determining
the thermodynamic properties of the liquid state, as well as the kinetics of phase transformations.
For example, liquid silicon displays many of the hallmark anomalies found in water, because
the covalent bonds in liquid silicon lead to tetrahedral structures, analogous to hydrogen-bonding
in water.
Due to the importance of silicon to the semiconductor industry, and technology as a whole, $l$-Si
is well-characterized.
However, past research has mainly focused on the structure and thermodynamics of liquid silicon ($l$-Si).
In the solid state at ambient conditions, silicon is a covalently-bonded semiconductor in the diamond lattice,
and upon melting it undergoes a semiconductor-to-metal transition~\cite{Sugino:PRL:1995,Alfe:PRB:2003,McMillan:2005ij,SCANNature}.
Early simulations indeed predicted that $l$-Si is metallic, in agreement with experiments,
but they also uncovered a non-negligible
fraction of covalent bonds that persist in the liquid state~\cite{Stich:PRB:1991,Okada:2012fb,Stich:PRL:1996,Stich:PRL:1989,Ashwin:2004oq,SCANNature,SiPRB,Remsing_2018}. 
The existence of these remnants of the solid phase were
later confirmed through a combination of computer simulations and Compton scattering experiments~\cite{Okada:2012fb}.
The formation of covalent bonds in the disordered liquid state indicates the presence of a competition
between metallic and covalent interactions in silicon.
Indeed, a quantitative description of the solid phases of Si necessitates an accurate model
for the balance of metallic and covalent interatomic interactions~\cite{SCANNature}.
This competition between metallic and covalent interactions is also predicted to underlie
a metallic-to-semimetallic liquid-liquid phase transition in silicon~\cite{Ashwin:2004oq,Ganesh:2009om,Beye:2010oq,Sastry:2010kl,Remsing_2018}. 
In this case, a high density metallic $l$-Si that is dominated by metallic bonding can transition to a low density
semimetallic liquid, in which the interatomic interactions are predominantly covalent bonds, albeit transient ones
that readily break and reform in response to thermal fluctuations~\cite{Ashwin:2004oq,Ganesh:2009om,Beye:2010oq,Sastry:2010kl,Remsing_2018}. 
Despite these significant investigations into the structure and thermodynamics of silicon,
quantification of the lifetimes of the transient covalent bonds in $l$-Si is lacking.
To address this issue, we use ab initio molecular dynamics (AIMD) simulations to characterize covalent bonding kinetics
in $l$-Si, which serves as a prototypical liquid-state system with dynamic covalent bonds.
After discussing simulation details in the next section, we quantify the structure of
transient covalent bonds in $l$-Si and present a geometric covalent bond definition. 
We then use this definition to quantify the dynamics of covalent bonding in $l$-Si
and conclude with a discussion of future directions.
%

\begin{figure*}[tb]
\begin{center}
\includegraphics[width=0.95\textwidth]{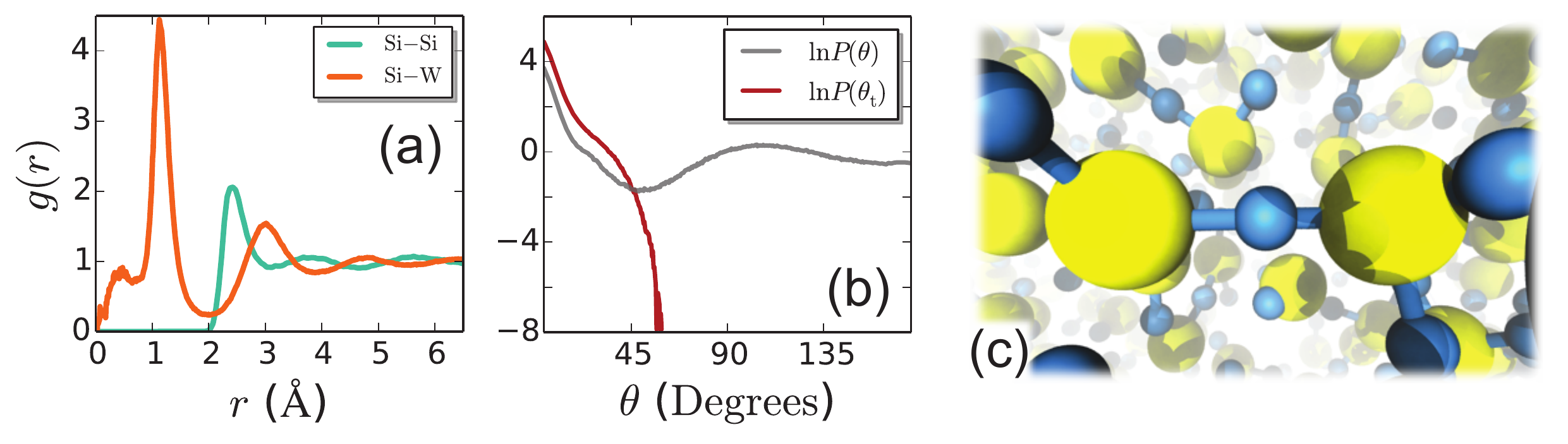}
\end{center}
\caption
{
(a) Pair distribution functions, $g(r)$, for correlations between silicon nuclei (Si-Si) and
between silicon atoms and maximally localized Wannier function centers (Si-W).
(b) The natural logarithms of the probability distributions of the Si-Si-W angle ($\theta$) and the Si-Si-W angle for triplets involving
the additional constraint that the Si-W distance is within the largest peak in the Si-W $g(r)$,
0.75~\AA$< r_{\rm Si-W} < 1.75$~\AA \ ($\theta_t$). The probability distributions are normalized
such that they equal unity for a uniform distribution.
(c) Snapshot of a Si-Si covalent bond in liquid Si satisfied the geometric definition proposed here.
Si atoms are shown in yellow and maximally localized Wannier function centers (W) are shown in blue.
}
\label{fig:struct}
\end{figure*}

\section{Simulation Details}

We simulated liquid Si at $T=1800$~K using the CP2K software package
following our previous work~\cite{Remsing_2018}.
The energies and forces in the MD simulations were evaluated using the \emph{QUICKSTEP} module~\cite{VandeVondele2005,VandeVondele2007},
which employs basis sets of Gaussian-type orbitals and plane waves for the electron density.
We used pseudopotentials, to represent the core electrons, and basis sets parameterized by
Godecker-Teter-Hutter (GTH):
GTH-PADE and GTH double-$\zeta$, single polarization (GTH-DZVP), respectively~\cite{Goedecker1996,VandeVondele2007}.
We explicitly treated the valence electrons of the 216 Si atoms using the strongly-constrained
and appropriately normed (SCAN) meta-generalized gradient approximation (meta-GGA)
density functional~\cite{SCAN,SCANNature}, as implemented in \texttt{LIBXC} version 4.0.1~\cite{LEHTOLA20181,MARQUES20122272},
with a plane wave cutoff of 650~Ry.
Initial configurations were taken from extensively equilibrated
simulations performed in earlier work~\cite{SiPRB,Remsing_2018}.
The systems were then further equilibrated at a constant temperature of $T=1800$~K,
maintained using the canonical velocity rescaling thermostat~\cite{Bussi:JCP:2007}.
Dynamic properties were computed from simulations in the microcanonical ensemble
using a timestep of 0.5~fs.
%

\section{Covalent Bond Structure in liquid Silicon}

We characterize the covalent bond structure in $l$-Si through the calculation
of maximally localized Wannier functions (MLWFs) and their centers (MLWFCs)~\cite{RevModPhys.84.1419}. 
MLWFs, in essence, can act as analogs of molecular orbitals for periodic systems, such as crystalline
and amorphous solids and liquids, and provide a useful, local picture of chemical bonding~\cite{Silvestrelli_SSC_1998,RevModPhys.84.1419}.
We use the MLWFCs to represent the position of electrons, as done previously~\cite{Silvestrelli_SSC_1998,RevModPhys.84.1419,Okada:2012fb,Remsing:JPCB:2019,Remsing:MolPhys:2018,Remsing:PRL:2020},
in order to quantify electron-nuclei correlations
and develop a geometric criterion defining the existence of a covalent bond in an atomic configuration.
We note that although the MLWFs themselves are not unique, the MLWFCs are invariant
with respect to the choice of gauge within a lattice vector, which is a time-independent constant
for simulations in constant volume ensembles, like the microcanonical ensemble used here~\cite{blount1962formalisms,RevModPhys.84.1419}. 
Thus, the MLWFCs can be used to define single bonds in each configuration.
We assume that covalent (single) bonds can be accurately defined by considering two- and three-body
correlations among Si nuclei and MLWFCs (W). 
The relevant two-body correlations are encoded in pair distribution functions,
$g(r)$, involving Si nuclei and MLWFCs.
Three-body correlations are captured by the probability distributions $P(\theta)$,
where $\theta$ is the Si-Si-MLWFC angle formed by an Si atom, its nearest neighbor Si atom,
and a MLWFC.
Both sets of distribution functions are shown in Figure~\ref{fig:struct}.
The pair distribution function for Si-Si and Si-MLWFC correlations, shown in Fig.~\ref{fig:struct}a,
are consistent with the formation of covalent bonds between Si atoms.
The first peak in the Si-W $g(r)$ is located halfway to the first Si-Si peak, and subsequent peaks
in the two correlation functions are out of phase.
Note that the small peak in the Si-W $g(r)$ at distances less than roughly 1~\AA \ is
consistent with the existence of non-bonded, lone pair electrons, visually depicted in Figures~\ref{fig:rf}a-c.
Triplet correlations, as quantified by $P(\theta)$, show that a significant fraction of nearest-neighbor MLWFCs are consistent
with covalent bonding, evidenced by the peak in $P(\theta)$ as $\theta$ approaches zero (Fig.~\ref{fig:struct}b), consistent
with linear Si-MLWFC-Si arrangements.
Large values of $\theta$ correspond to Si-Si-W triplets not involved in covalent bonds.
We additionally consider the possibility that lone pair-like MLWFCs may also appear at low values of $\theta$.
Thus, we place the additional constraint on the Si-Si-W angle that the Si-W distance must be between 0.75~\AA \
and 1.75~\AA \ to avoid counting lone pair MLWFCs, resulting in the angle $\theta_t$.
This more tightly-constrained angle removes contributions from lone pairs to the angular
distribution $P(\theta_t)$, Fig.~\ref{fig:struct}b.
With this additional constraint, we suggest a reasonable geometric definition of a Si-Si covalent bond
corresponds to a Si-Si distance less than 3~\AA \ and $\theta_t<30\degree$.
One such covalent bond is highlighted in Fig.~\ref{fig:struct}c.
%

\begin{figure}[tb]
\begin{center}
\includegraphics[width=0.5\textwidth]{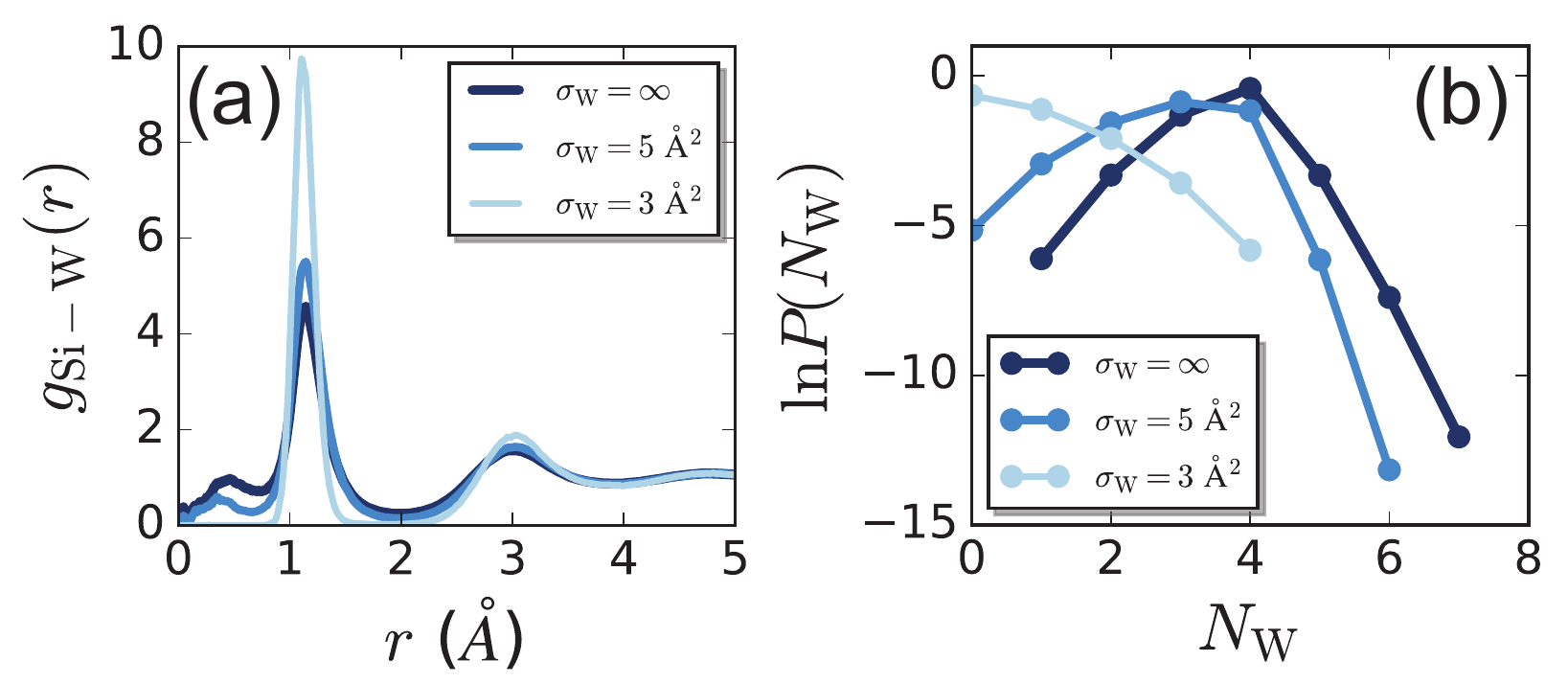}
\end{center}
\caption
{
(a) Pair distribution functions, $g(r)$, for correlations between silicon nuclei
and maximally localized Wannier function centers (Si-W)
for indicated values of the cutoff for the spread of the MLWFs, $\sigma_{\rm W}$,
above which we exclude MLWFCs from the calculation. 
(b) Probability distributions, $P(N_{\rm W})$, of the number of MLWFCs, $N_{\rm W}$,
within a distance of 1.75~\AA \ of a central Si atom for the same values of $\sigma_{\rm W}$.
}
\label{fig:spread}
\end{figure}

%
In the snapshot shown in Fig.~\ref{fig:struct}c, as well as those in Figs.~\ref{fig:rf}a-c,
one can observe a range of bonding and coordination environments. 
In particular, the number of MLWFCs associated with a single Si atom varies,
and, similarly, the number of Si atoms associated with a MLWFC also varies. 
Characterizing the fluctuations in Si and MLWFC coordination structures
quantifies the probability of forming lone pairs, covalent bond pairs, and metallic or diffuse pairs~\cite{Okada:2012fb,Okada_2015}. 
However, positional correlations alone do not suffice to characterize the nature of the MLWFs, as discussed
above in the context of angular correlations. 
Lone pairs and covalent bond MLWFs are high localized in space, such that their spreads are small,
such that bonded MLWFs are generally more localized than lone pair MLWFs~\cite{Okada:2012fb,Okada_2015}.
In contrast, the spreads of metallic or diffuse MLWFs are large, corresponding to delocalized pairs.
Therefore, we quantify the coordination structure of Si atoms and MLWFCs through distances,
as well as a range of MLWF spreads. 
We introduce a MLWF spread cutoff, $\sigma_{\rm W}$, such that MLWFs with spreads above this value are not included
in averages; $\sigma_{\rm W}=\infty$ indicates that all MLWFCs are included in the calculations.
We start by examining the impact of $\sigma_{\rm W}$ on the Si-MLWFC pair distribution function,
$g_{\rm Si-W}(r)$, shown in Fig.~\ref{fig:spread}a.
Introducing a finite $\sigma_{\rm W}$ removes the contribution of diffuse MLWFs from the pair
distribution function.
A spread cutoff of $5$~\AA$^2$ reduces the peak at small distances and slightly increases
the first major peak, as well as the second peak. 
Further decreasing $\sigma_{\rm W}$ to $3$~\AA$^2$ removes diffuse and nearly all lone pairs,
evidence by an absence of a peak for $r<1$~\AA. 
The MLWFs remaining in the calculation are predominantly covalently bonded pairs, 
and the intensity of the first major peak in $g_{\rm Si-W}(r)$, corresponding to bonded pairs,
increasingly significantly.
The second peak increases slightly as well. 
Therefore, $\sigma_{\rm W}$ can be tuned to remove diffuse and lone pair MLWFCs,
and this tuning has the mainly impacts the structure of the first coordination shell, $r<1.75$~\AA.
The probabilities of observing $N_{\rm W}$ MLWFCs in the first coordination shell of an Si atom
for various $\sigma_{\rm W}$ are shown in Fig.~\ref{fig:spread}b.
When all MLWFCs are included, $N_{\rm W}$ ranges from 1-7, and a maximum is observed at
$N_{\rm W}=4$, as expect from the $sp^3$ hybridization of the Si atoms.
As the cutoff is decreased, $P(N_{\rm W})$ shifts toward lower values, as delocalized MLWFs
are removed from consideration.
For $\sigma_{\rm W}=5$~\AA$^2$, the maximum shifts to $N_{\rm W}=3$, with similar probabilities
at $N_{\rm W}=2$ and $N_{\rm W}=4$.
The appearance of finite probability for $N_{\rm W}=0$ is consistent with Si atoms that have all their
electrons in diffuse, metallic states.
Further reduction of $\sigma_{\rm W}$ to $3$~\AA$^2$, which limits
the set of MLWFCs almost entirely to covalently bonded pairs, shifts the maximum of $P(N_{\rm W})$
to zero, consistent with a small fraction (roughly 30 percent) of Si atoms involved in covalent bonds. 
In this limit, $P(N_{\rm W})$ essentially corresponds to probability of an Si atom
having $N_{\rm W}$ covalent bonds.
This spans zero to four bonds per Si atom, and monotonically decreases with $N_{\rm W}$,
indicating that Si atoms fully coordinated by covalent bonds are less probable than partially
and non-bonded Si atoms.
This is consistent with the metallic nature of liquid Si.
%

\begin{figure*}[tb]
\begin{center}
\includegraphics[width=0.99\textwidth]{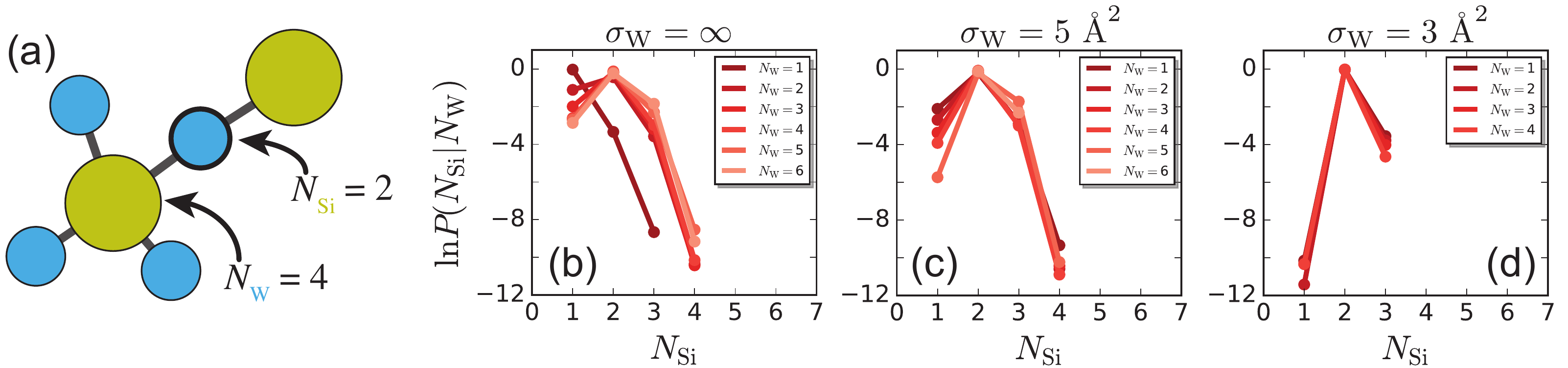}
\end{center}
\caption
{
(a) Schematic diagram indicating the examples of $N_{\rm W}$ and the number of Si atoms
around a MLWFC that is in the coordination shell of a Si atom, $N_{\rm Si}$.
Si atoms are colored yellow and MLWFCs are colored blue.
Arrows point toward the Si and MLWFC to which the indicated $N_{\rm W}$ and $N_{\rm Si}$ refer.
(b-d) Conditional probability distributions, $P(N_{\rm Si} | N_{\rm W})$, of the number of 
Si atoms within a distance of 1.75~\AA \ of a MLWFC, where the MLWFC is
part of a Si coordination shell composed of $N_{\rm W}$ MLWFCs,
for the indicated values of the MLWFC spread cutoff, $\sigma_{\rm W}$.
}
\label{fig:joint}
\end{figure*}

%
We now turn our attention to the coordination structure of the MLWFCs, particularly
those that are in the first coordination of a Si atom.
To quantify this, we compute the joint probability distribution, $P(N_{\rm Si}, N_{\rm W})$,
corresponding to the probability that a central Si atom is coordinated by $N_{\rm W}$ MLWFCs
and one of those coordinated MLWFCs is in turn coordinated by $N_{\rm Si}$ Si atoms
(including the central Si atom). 
A schematic for $N_{\rm W}=4$ and $N_{\rm Si}=2$ for one MLWFC is shown in Fig.~\ref{fig:joint}a. 
For ease of visualization, we focus on the conditional probability,
\begin{equation}
P(N_{\rm Si} | N_{\rm W})=\frac{P(N_{\rm Si}, N_{\rm W})}{P(N_{\rm W})},
\end{equation}
physically corresponding to the probability that a MLWFC in the coordination shell of a central Si
atom is coordinated by $N_{\rm Si}$ Si atoms, given that the central Si has $N_{\rm W}$ MLWFCs
in its first coordination shell. 
These distributions are shown in Fig.~\ref{fig:joint}b-c for varying $\sigma_{\rm W}$.
When all MLWFCs are included in $P(N_{\rm Si} | N_{\rm W})$, the conditional probability
has a maximum at $N_{\rm Si}=2$ for all values of $N_{\rm W}>1$. 
This maximum corresponds to a covalent bond between the central Si and
a neighboring Si atom.
Reducing $N_{\rm W}$ from 6 to 2 primarily impacts the probably of observing singly coordinated
MLWFCs (to the central Si), which increases as $N_{\rm W}$ is lowered.
For $N_{\rm W}=1$, $P(N_{\rm Si} | N_{\rm W})$
differs significantly from the rest, with $N_{\rm Si}=1$ being most probable,
indicating that the MLWFC is most likely a diffuse or lone pair. 
Introducing a finite MLWF spread cutoff of $\sigma_{\rm W}=5$~\AA$^2$
removes diffuse pairs and results in distributions that are similar for all $N_{\rm Si}$,
except $N_{\rm Si}=1$.
As $N_{\rm W}$ is decreased, the probability of observing singly-coordinated, lone pair
MLWFCs increases.
Further reduction of $\sigma_{\rm W}$ to $3$~\AA$^2$ results in sharp
$P(N_{\rm Si} | N_{\rm W})$ distributions that are nearly independent of $N_{\rm W}$.
These distributions span $1\le N_{\rm Si}\le3$, with a large maximum at $N_{\rm Si}=2$,
corresponding to covalent bond MLWFCs.
We note that there is a small probability for lone pairs when $\sigma_{\rm W}=3$~\AA$^2$,
but lowering the spread cutoff further will increasingly remove these MLWFCs
and select only covalently bonded pairs~\cite{Okada:2012fb,Okada_2015}.
In summary, the local (bonding) coordination structure of Si atoms significantly fluctuates
and involves diffuse metallic, lone pair, and covalently-bonded states.
We have characterized these states, and the use of a MLWF spread cutoff, $\sigma_{\rm W}$,
can be used to systematically tune the involvement of these states when necessary.
Alternatively, Si-Si-W angular correlations, in addition to Si-Si and Si-W distances, used
in the above-described geometric definition of covalent bonds can be used to uncover
covalently-bonded MLWFCs in a similar manner, because linear Si-W-Si structures
are consistent with $N_{\rm Si}=2$ covalently-bonded pairs.
%

\section{Covalent Bond Dynamics in liquid Silicon}

\begin{figure}[tbh]
\begin{center}
\includegraphics[width=0.42\textwidth]{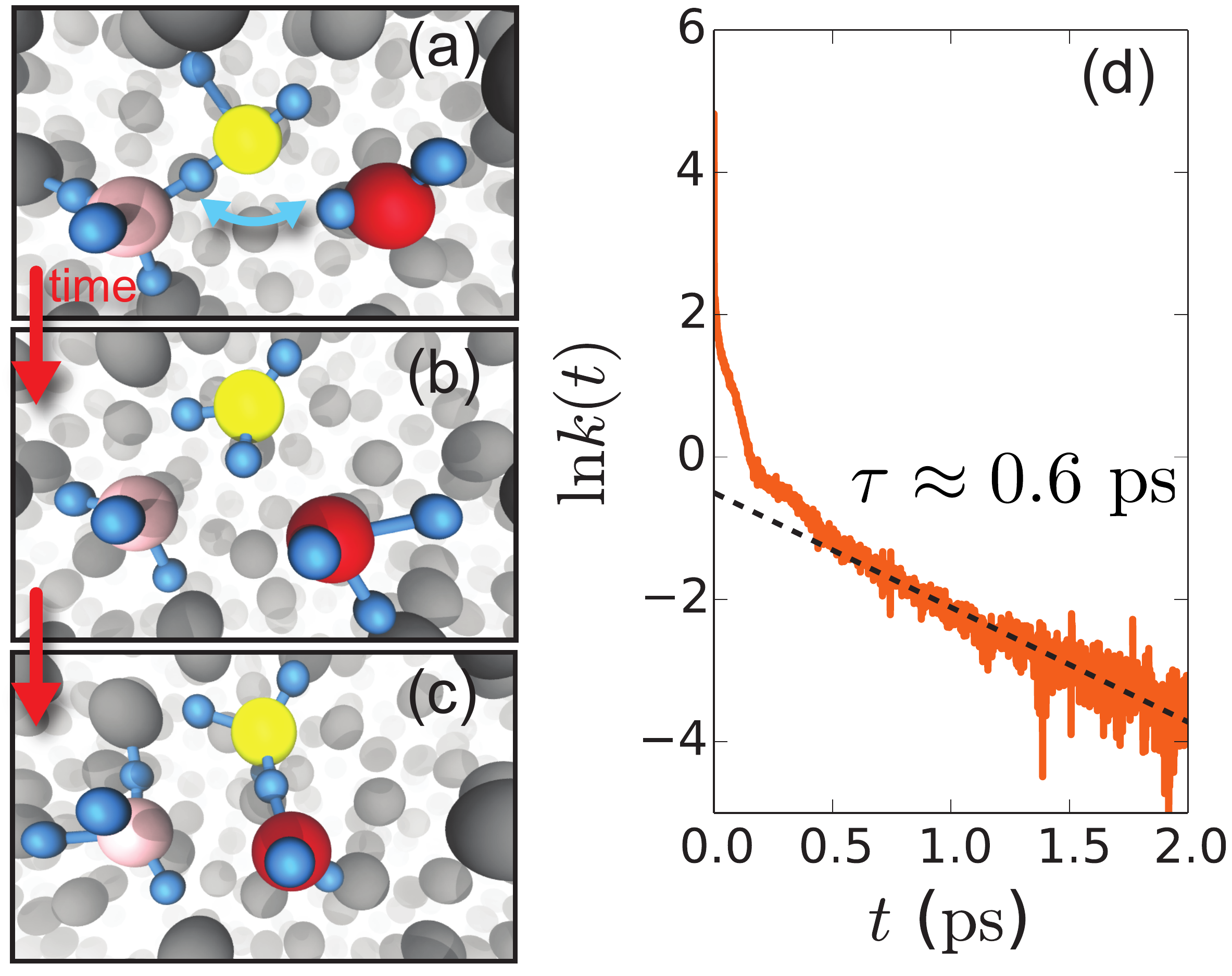}
\end{center}
\caption
{
(a-c) Snapshots of covalent bond breakage and reformation in liquid silicon, with three Si atoms involved
in the events colored yellow, pink, and red; all other Si atoms are colored grey.
Maximally localized Wannier function centers (MLWFCs) associated with the colored silicon atoms are shown in blue.
(a) Initially, the central (yellow) Si is bonded to the pink Si atom on the left, as indicated by the lines connecting
the Si atoms through a bridging MLWFC.
(b) At some time later, the bond breaks due to thermal fluctuations, and the MLWFC does not connect to Si atoms.
(c) A new bond then forms between the central Si atom and the rightmost (red) colored Si atom. 
The left (pink) Si has also formed a new bond on this timescale.
(d) The kinetics of bond covalent bond breakage is quantified by the reactive flux correlation function, $k(t)$,
and fitting its long time behavior (dashed line) results in a covalent bond lifetime of $\tau\approx 0.6$~ps.
}
\label{fig:rf}
\end{figure}

Our simulations suggest that covalent bonds in $l$-Si rapidly break and reform on sub-picosecond timescales.
Figures~\ref{fig:rf}a-c highlight one such covalent bond breakage and reformation event.
There, we show the time evolution of the MLWFCs (blue spheres)
for the atoms involved in the highlighted covalent bond exchange.
Initially (a), the central yellow Si atom is bonded to the left pink Si atom; a linear Si-MLWFC-Si
structure indicates a single covalent bond expected from the sp$^3$ hybridization of Si.
This bond breaks at a later time (b) due to thermal fluctuations, before the MLWFC of the central Si rotates
and forms a new bond with the rightmost red Si (c). 
Using the geometric definition of a covalent bond described in the previous section,
we are able to quantify the kinetics of Si-Si bond breakage,
in a manner analogous to conventional approaches to characterizing hydrogen bond dynamics in water~\cite{Chandler_1978,Luzar_2000,LuzarChandler},
and, more recently, halogen bond dynamics in solid and liquid chlorine~\cite{Remsing:JPCB:2019, Remsing:PRL:2020}.
To do so, we define an indicator function, $h(t)$, which is equal to one when a covalent bond is present and zero otherwise.
We quantify covalent bond kinetics through the reactive flux correlation function~\cite{Chandler_1978}
\begin{equation}
k(t)=-\frac{d C(t)}{dt}=-\frac{\avg{\dot{h}(0)\brac{1-h(t)}}}{\avg{h}},
\end{equation}
where $C(t)$ is the time correlation function (TCF) characterizing covalent bond lifetimes,
\begin{equation}
C(t)=\frac{\avg{h(t) h(0)}}{\avg{h}}.
\end{equation}
%

%
The reactive flux correlation function, $k(t)$, is shown in Figure~\ref{fig:rf}d.
At short times, vibrational and librational motion manifest the non-trivial, transient behavior
on scales less than roughly 0.3~ps. 
Beyond this time period, $k(t)$ decays in a manner consistent with first-order kinetics,
$k(t)\sim \tau^{-1} \exp(t/\tau)$, where $\tau\approx 0.6$~ps is the covalent bond lifetime
estimated from fitting to the long-time behavior of $k(t)$, shown as a dashed line
in Fig.~\ref{fig:rf}d.
The covalent bond lifetime estimated above is similar to other significant timescales in $l$-Si. 
Orientational correlations, the velocity autocorrelation function,
and self-intermediate scattering functions all decay on timescales
similar to the covalent bond lifetime~\cite{Remsing_2018,SiPRB}. 
The agreement among various structural relaxation times and the covalent bond lifetime suggests that
covalent bond breakage is a limiting step for structural relaxation. 
This may be expected from the strength of covalent bonds, as well as their propensity to create
local tetrahedral order in the liquid state. 
%

\section{Conclusions}	

In this work, we have presented an approach to estimate the lifetimes of transient covalent bonds
in condensed phase systems with an application to liquid silicon. 
The general strategy utilizes a maximally localized Wannier function (MLWF) approach to chemical bonding,
such that the existence of a covalent bond can be defined using two-body Si-Si correlations
and three-body correlations involving two Si nuclei and the center of a MLWF. 
With this geometric, ab initio definition of a covalent bond, we can estimate covalent bond lifetimes
in the liquid state from equilibrium simulations using the reactive flux formalism.
For liquid silicon at 1800~K and ambient pressure, we estimate a covalent bond lifetime of $\tau\approx 0.6$~ps.
We note, however, that the generic concept of monitoring bond dynamics using a geometric, electronic structure-based
bond definition is not limited to equilibrium and could be used to monitor covalent bond dynamics in
melting or freezing processes at the focus of
laster melting experiments~\cite{Beye:2010oq,Stiffler:PRL:1988,Sanders:JApplPhys:1999}, for example.
The covalent bond lifetime in liquid silicon may be measured using time-dependent scattering measurements,
such a time-dependent Compton scattering. 
Compton scattering has been used to shed light on the average bonding properties of metallic liquids~\cite{Okada:2012fb,Okada_2015}.
Extensions of this technique to the time domain are expected to uncover similar information about the average
dynamic properties of transient covalent bonds, like the lifetime at the focus of this work~\cite{Wagner_2010,Kemper_2013,Grosser_2017}.

Finally, we note that the covalent bond lifetime in $l$-Si is similar to bond lifetimes in other conventional
liquids with directional attractive interactions.
The hydrogen bond lifetime in liquid water is on the picosecond timescale~\cite{Luzar_2000,LuzarChandler},
as is the halogen bond lifetime in liquid Cl$_2$~\cite{Remsing:JPCB:2019}.
Despite the vast differences in these liquids, their directional bonds all share a common thread:
the strength of the isolated bond, which is significantly weakened in the condensed phase, is on the order of $10\kT$
at the temperature of the respective liquids~\cite{bryant2012water}.
Thus, thermal fluctuations in each of these different liquids are large enough
to cause the directional attractions to exist only fleetingly,
highlighting qualitative similarities of directional bond dynamics in the liquid state.

\begin{acknowledgements}
This work was supported as part of the
Center for Complex Materials from First Principles (CCM), an Energy Frontier
Research Center funded by the U.S. Department of Energy, Office of Science, Basic Energy
Sciences under Award \#DE-SC0012575.
Computational resources were supported in part by the National Science Foundation
through major research instrumentation grant number 1625061
and by the US ARL under contract number W911NF-16-2-0189.
\end{acknowledgements}

\bibliography{Silicon}

\begin{thebibliography}{53}%
\makeatletter
\providecommand \@ifxundefined [1]{%
 \@ifx{#1\undefined}
}%
\providecommand \@ifnum [1]{%
 \ifnum #1\expandafter \@firstoftwo
 \else \expandafter \@secondoftwo
 \fi
}%
\providecommand \@ifx [1]{%
 \ifx #1\expandafter \@firstoftwo
 \else \expandafter \@secondoftwo
 \fi
}%
\providecommand \natexlab [1]{#1}%
\providecommand \enquote  [1]{``#1''}%
\providecommand \bibnamefont  [1]{#1}%
\providecommand \bibfnamefont [1]{#1}%
\providecommand \citenamefont [1]{#1}%
\providecommand \href@noop [0]{\@secondoftwo}%
\providecommand \href [0]{\begingroup \@sanitize@url \@href}%
\providecommand \@href[1]{\@@startlink{#1}\@@href}%
\providecommand \@@href[1]{\endgroup#1\@@endlink}%
\providecommand \@sanitize@url [0]{\catcode `\\12\catcode `\$12\catcode
  `\&12\catcode `\#12\catcode `\^12\catcode `\_12\catcode `\%12\relax}%
\providecommand \@@startlink[1]{}%
\providecommand \@@endlink[0]{}%
\providecommand \url  [0]{\begingroup\@sanitize@url \@url }%
\providecommand \@url [1]{\endgroup\@href {#1}{\urlprefix }}%
\providecommand \urlprefix  [0]{URL }%
\providecommand \Eprint [0]{\href }%
\providecommand \doibase [0]{http://dx.doi.org/}%
\providecommand \selectlanguage [0]{\@gobble}%
\providecommand \bibinfo  [0]{\@secondoftwo}%
\providecommand \bibfield  [0]{\@secondoftwo}%
\providecommand \translation [1]{[#1]}%
\providecommand \BibitemOpen [0]{}%
\providecommand \bibitemStop [0]{}%
\providecommand \bibitemNoStop [0]{.\EOS\space}%
\providecommand \EOS [0]{\spacefactor3000\relax}%
\providecommand \BibitemShut  [1]{\csname bibitem#1\endcsname}%
\let\auto@bib@innerbib\@empty
\bibitem [{\citenamefont {\v{S}tich}, \citenamefont {Car},\ and\ \citenamefont
  {Parrinello}(1991)}]{Stich:PRB:1991}%
  \BibitemOpen
  \bibfield  {author} {\bibinfo {author} {\bibfnamefont {I.}~\bibnamefont
  {\v{S}tich}}, \bibinfo {author} {\bibfnamefont {R.}~\bibnamefont {Car}}, \
  and\ \bibinfo {author} {\bibfnamefont {M.}~\bibnamefont {Parrinello}},\
  }\href@noop {} {\bibfield  {journal} {\bibinfo  {journal} {Phys. Rev. B}\
  }\textbf {\bibinfo {volume} {44}},\ \bibinfo {pages} {4262} (\bibinfo {year}
  {1991})}\BibitemShut {NoStop}%
\bibitem [{\citenamefont {Okada}\ \emph {et~al.}(2012)\citenamefont {Okada},
  \citenamefont {Sit}, \citenamefont {Watanabe}, \citenamefont {Wang},
  \citenamefont {Barbiellini}, \citenamefont {Ishikawa}, \citenamefont {Itou},
  \citenamefont {Sakurai}, \citenamefont {Bansil}, \citenamefont {Ishikawa},
  \citenamefont {Hamaishi}, \citenamefont {Masaki}, \citenamefont {Paradis},
  \citenamefont {Kimura}, \citenamefont {Ishikawa},\ and\ \citenamefont
  {Nanao}}]{Okada:2012fb}%
  \BibitemOpen
  \bibfield  {author} {\bibinfo {author} {\bibfnamefont {J.~T.}\ \bibnamefont
  {Okada}}, \bibinfo {author} {\bibfnamefont {P.~H.-L.}\ \bibnamefont {Sit}},
  \bibinfo {author} {\bibfnamefont {Y.}~\bibnamefont {Watanabe}}, \bibinfo
  {author} {\bibfnamefont {Y.~J.}\ \bibnamefont {Wang}}, \bibinfo {author}
  {\bibfnamefont {B.}~\bibnamefont {Barbiellini}}, \bibinfo {author}
  {\bibfnamefont {T.}~\bibnamefont {Ishikawa}}, \bibinfo {author}
  {\bibfnamefont {M.}~\bibnamefont {Itou}}, \bibinfo {author} {\bibfnamefont
  {Y.}~\bibnamefont {Sakurai}}, \bibinfo {author} {\bibfnamefont
  {A.}~\bibnamefont {Bansil}}, \bibinfo {author} {\bibfnamefont
  {R.}~\bibnamefont {Ishikawa}}, \bibinfo {author} {\bibfnamefont
  {M.}~\bibnamefont {Hamaishi}}, \bibinfo {author} {\bibfnamefont
  {T.}~\bibnamefont {Masaki}}, \bibinfo {author} {\bibfnamefont {P.-F.}\
  \bibnamefont {Paradis}}, \bibinfo {author} {\bibfnamefont {K.}~\bibnamefont
  {Kimura}}, \bibinfo {author} {\bibfnamefont {T.}~\bibnamefont {Ishikawa}}, \
  and\ \bibinfo {author} {\bibfnamefont {S.}~\bibnamefont {Nanao}},\
  }\href@noop {} {\bibfield  {journal} {\bibinfo  {journal} {Phys. Rev. Lett.}\
  }\textbf {\bibinfo {volume} {108}},\ \bibinfo {pages} {067402} (\bibinfo
  {year} {2012})}\BibitemShut {NoStop}%
\bibitem [{\citenamefont {\v{S}tich}, \citenamefont {Parrinello},\ and\
  \citenamefont {Holender}(1996)}]{Stich:PRL:1996}%
  \BibitemOpen
  \bibfield  {author} {\bibinfo {author} {\bibfnamefont {I.}~\bibnamefont
  {\v{S}tich}}, \bibinfo {author} {\bibfnamefont {M.}~\bibnamefont
  {Parrinello}}, \ and\ \bibinfo {author} {\bibfnamefont {J.~M.}\ \bibnamefont
  {Holender}},\ }\href@noop {} {\bibfield  {journal} {\bibinfo  {journal}
  {Phys. Rev. Lett.}\ }\textbf {\bibinfo {volume} {76}},\ \bibinfo {pages}
  {2077} (\bibinfo {year} {1996})}\BibitemShut {NoStop}%
\bibitem [{\citenamefont {\v{S}tich}, \citenamefont {Car},\ and\ \citenamefont
  {Parrinello}(1989)}]{Stich:PRL:1989}%
  \BibitemOpen
  \bibfield  {author} {\bibinfo {author} {\bibfnamefont {I.}~\bibnamefont
  {\v{S}tich}}, \bibinfo {author} {\bibfnamefont {R.}~\bibnamefont {Car}}, \
  and\ \bibinfo {author} {\bibfnamefont {M.}~\bibnamefont {Parrinello}},\
  }\href@noop {} {\bibfield  {journal} {\bibinfo  {journal} {Phys. Rev. Lett.}\
  }\textbf {\bibinfo {volume} {63}},\ \bibinfo {pages} {2240} (\bibinfo {year}
  {1989})}\BibitemShut {NoStop}%
\bibitem [{\citenamefont {Ashwin}, \citenamefont {Waghmare},\ and\
  \citenamefont {Sastry}(2004)}]{Ashwin:2004oq}%
  \BibitemOpen
  \bibfield  {author} {\bibinfo {author} {\bibfnamefont {S.~S.}\ \bibnamefont
  {Ashwin}}, \bibinfo {author} {\bibfnamefont {U.~V.}\ \bibnamefont
  {Waghmare}}, \ and\ \bibinfo {author} {\bibfnamefont {S.}~\bibnamefont
  {Sastry}},\ }\href {\doibase 10.1103/PhysRevLett.92.175701} {\bibfield
  {journal} {\bibinfo  {journal} {Phys Rev Lett}\ }\textbf {\bibinfo {volume}
  {92}},\ \bibinfo {pages} {175701} (\bibinfo {year} {2004})}\BibitemShut
  {NoStop}%
\bibitem [{\citenamefont {Sun}\ \emph {et~al.}(2016)\citenamefont {Sun},
  \citenamefont {Remsing}, \citenamefont {Zhang}, \citenamefont {Sun},
  \citenamefont {Ruzsinszky}, \citenamefont {Peng}, \citenamefont {Yang},
  \citenamefont {Paul}, \citenamefont {Waghmare}, \citenamefont {Wu},
  \citenamefont {Klein},\ and\ \citenamefont {Perdew}}]{SCANNature}%
  \BibitemOpen
  \bibfield  {author} {\bibinfo {author} {\bibfnamefont {J.}~\bibnamefont
  {Sun}}, \bibinfo {author} {\bibfnamefont {R.~C.}\ \bibnamefont {Remsing}},
  \bibinfo {author} {\bibfnamefont {Y.}~\bibnamefont {Zhang}}, \bibinfo
  {author} {\bibfnamefont {Z.}~\bibnamefont {Sun}}, \bibinfo {author}
  {\bibfnamefont {A.}~\bibnamefont {Ruzsinszky}}, \bibinfo {author}
  {\bibfnamefont {H.}~\bibnamefont {Peng}}, \bibinfo {author} {\bibfnamefont
  {Z.}~\bibnamefont {Yang}}, \bibinfo {author} {\bibfnamefont {A.}~\bibnamefont
  {Paul}}, \bibinfo {author} {\bibfnamefont {U.}~\bibnamefont {Waghmare}},
  \bibinfo {author} {\bibfnamefont {X.}~\bibnamefont {Wu}}, \bibinfo {author}
  {\bibfnamefont {M.~L.}\ \bibnamefont {Klein}}, \ and\ \bibinfo {author}
  {\bibfnamefont {J.~P.}\ \bibnamefont {Perdew}},\ }\href {\doibase
  10.1038/nchem.2535} {\bibfield  {journal} {\bibinfo  {journal} {Nat Chem}\
  }\textbf {\bibinfo {volume} {8}},\ \bibinfo {pages} {831} (\bibinfo {year}
  {2016})}\BibitemShut {NoStop}%
\bibitem [{\citenamefont {Remsing}, \citenamefont {Klein},\ and\ \citenamefont
  {Sun}(2017)}]{SiPRB}%
  \BibitemOpen
  \bibfield  {author} {\bibinfo {author} {\bibfnamefont {R.~C.}\ \bibnamefont
  {Remsing}}, \bibinfo {author} {\bibfnamefont {M.~L.}\ \bibnamefont {Klein}},
  \ and\ \bibinfo {author} {\bibfnamefont {J.}~\bibnamefont {Sun}},\
  }\href@noop {} {\bibfield  {journal} {\bibinfo  {journal} {Phys. Rev. B}\
  }\textbf {\bibinfo {volume} {96}},\ \bibinfo {pages} {024203} (\bibinfo
  {year} {2017})}\BibitemShut {NoStop}%
\bibitem [{\citenamefont {Remsing}, \citenamefont {Klein},\ and\ \citenamefont
  {Sun}(2018)}]{Remsing_2018}%
  \BibitemOpen
  \bibfield  {author} {\bibinfo {author} {\bibfnamefont {R.~C.}\ \bibnamefont
  {Remsing}}, \bibinfo {author} {\bibfnamefont {M.~L.}\ \bibnamefont {Klein}},
  \ and\ \bibinfo {author} {\bibfnamefont {J.}~\bibnamefont {Sun}},\
  }\href@noop {} {\bibfield  {journal} {\bibinfo  {journal} {Phys. Rev. B}\
  }\textbf {\bibinfo {volume} {97}},\ \bibinfo {pages} {140103(R)} (\bibinfo
  {year} {2018})}\BibitemShut {NoStop}%
\bibitem [{\citenamefont {Okada}\ \emph {et~al.}(2015)\citenamefont {Okada},
  \citenamefont {Sit}, \citenamefont {Watanabe}, \citenamefont {Barbiellini},
  \citenamefont {Ishikawa}, \citenamefont {Wang}, \citenamefont {Itou},
  \citenamefont {Sakurai}, \citenamefont {Bansil}, \citenamefont {Ishikawa},\
  and\ \citenamefont {et~al.}}]{Okada_2015}%
  \BibitemOpen
  \bibfield  {author} {\bibinfo {author} {\bibfnamefont {J.}~\bibnamefont
  {Okada}}, \bibinfo {author} {\bibfnamefont {P.-L.}\ \bibnamefont {Sit}},
  \bibinfo {author} {\bibfnamefont {Y.}~\bibnamefont {Watanabe}}, \bibinfo
  {author} {\bibfnamefont {B.}~\bibnamefont {Barbiellini}}, \bibinfo {author}
  {\bibfnamefont {T.}~\bibnamefont {Ishikawa}}, \bibinfo {author}
  {\bibfnamefont {Y.}~\bibnamefont {Wang}}, \bibinfo {author} {\bibfnamefont
  {M.}~\bibnamefont {Itou}}, \bibinfo {author} {\bibfnamefont {Y.}~\bibnamefont
  {Sakurai}}, \bibinfo {author} {\bibfnamefont {A.}~\bibnamefont {Bansil}},
  \bibinfo {author} {\bibfnamefont {R.}~\bibnamefont {Ishikawa}}, \ and\
  \bibinfo {author} {\bibnamefont {et~al.}},\ }\href@noop {} {\bibfield
  {journal} {\bibinfo  {journal} {Phys. Rev. Lett.}\ }\textbf {\bibinfo
  {volume} {114}},\ \bibinfo {pages} {177401} (\bibinfo {year}
  {2015})}\BibitemShut {NoStop}%
\bibitem [{\citenamefont {Gong}\ \emph {et~al.}(1993)\citenamefont {Gong},
  \citenamefont {Chiarotti}, \citenamefont {Parrinello},\ and\ \citenamefont
  {Tosatti}}]{Gong_EL_1993}%
  \BibitemOpen
  \bibfield  {author} {\bibinfo {author} {\bibfnamefont {X.~G.}\ \bibnamefont
  {Gong}}, \bibinfo {author} {\bibfnamefont {G.~L.}\ \bibnamefont {Chiarotti}},
  \bibinfo {author} {\bibfnamefont {M.}~\bibnamefont {Parrinello}}, \ and\
  \bibinfo {author} {\bibfnamefont {E.}~\bibnamefont {Tosatti}},\ }\href@noop
  {} {\bibfield  {journal} {\bibinfo  {journal} {Europhys. Lett.}\ }\textbf
  {\bibinfo {volume} {21}},\ \bibinfo {pages} {469} (\bibinfo {year}
  {1993})}\BibitemShut {NoStop}%
\bibitem [{\citenamefont {Rillo}\ \emph {et~al.}(2019)\citenamefont {Rillo},
  \citenamefont {Morales}, \citenamefont {Ceperley},\ and\ \citenamefont
  {Pierleoni}}]{Rillo_2019}%
  \BibitemOpen
  \bibfield  {author} {\bibinfo {author} {\bibfnamefont {G.}~\bibnamefont
  {Rillo}}, \bibinfo {author} {\bibfnamefont {M.~A.}\ \bibnamefont {Morales}},
  \bibinfo {author} {\bibfnamefont {D.~M.}\ \bibnamefont {Ceperley}}, \ and\
  \bibinfo {author} {\bibfnamefont {C.}~\bibnamefont {Pierleoni}},\ }\href
  {\doibase 10.1073/pnas.1818897116} {\bibfield  {journal} {\bibinfo  {journal}
  {Proceedings of the National Academy of Sciences}\ }\textbf {\bibinfo
  {volume} {116}},\ \bibinfo {pages} {9770} (\bibinfo {year}
  {2019})}\BibitemShut {NoStop}%
\bibitem [{\citenamefont {Morales}\ \emph {et~al.}(2013)\citenamefont
  {Morales}, \citenamefont {McMahon}, \citenamefont {Pierleoni},\ and\
  \citenamefont {Ceperley}}]{Morales_2013}%
  \BibitemOpen
  \bibfield  {author} {\bibinfo {author} {\bibfnamefont {M.~A.}\ \bibnamefont
  {Morales}}, \bibinfo {author} {\bibfnamefont {J.~M.}\ \bibnamefont
  {McMahon}}, \bibinfo {author} {\bibfnamefont {C.}~\bibnamefont {Pierleoni}},
  \ and\ \bibinfo {author} {\bibfnamefont {D.~M.}\ \bibnamefont {Ceperley}},\
  }\href {\doibase 10.1103/physrevlett.110.065702} {\bibfield  {journal}
  {\bibinfo  {journal} {Physical Review Letters}\ }\textbf {\bibinfo {volume}
  {110}} (\bibinfo {year} {2013}),\ 10.1103/physrevlett.110.065702}\BibitemShut
  {NoStop}%
\bibitem [{\citenamefont {Pierleoni}, \citenamefont {Holzmann},\ and\
  \citenamefont {Ceperley}(2017)}]{Pierleoni_2017}%
  \BibitemOpen
  \bibfield  {author} {\bibinfo {author} {\bibfnamefont {C.}~\bibnamefont
  {Pierleoni}}, \bibinfo {author} {\bibfnamefont {M.}~\bibnamefont {Holzmann}},
  \ and\ \bibinfo {author} {\bibfnamefont {D.~M.}\ \bibnamefont {Ceperley}},\
  }\href {\doibase 10.1002/ctpp.201700184} {\bibfield  {journal} {\bibinfo
  {journal} {Contributions to Plasma Physics}\ }\textbf {\bibinfo {volume}
  {58}},\ \bibinfo {pages} {99} (\bibinfo {year} {2017})}\BibitemShut {NoStop}%
\bibitem [{\citenamefont {Pierleoni}\ \emph {et~al.}(2016)\citenamefont
  {Pierleoni}, \citenamefont {Morales}, \citenamefont {Rillo}, \citenamefont
  {Holzmann},\ and\ \citenamefont {Ceperley}}]{Pierleoni_2016}%
  \BibitemOpen
  \bibfield  {author} {\bibinfo {author} {\bibfnamefont {C.}~\bibnamefont
  {Pierleoni}}, \bibinfo {author} {\bibfnamefont {M.~A.}\ \bibnamefont
  {Morales}}, \bibinfo {author} {\bibfnamefont {G.}~\bibnamefont {Rillo}},
  \bibinfo {author} {\bibfnamefont {M.}~\bibnamefont {Holzmann}}, \ and\
  \bibinfo {author} {\bibfnamefont {D.~M.}\ \bibnamefont {Ceperley}},\ }\href
  {\doibase 10.1073/pnas.1603853113} {\bibfield  {journal} {\bibinfo  {journal}
  {Proceedings of the National Academy of Sciences}\ }\textbf {\bibinfo
  {volume} {113}},\ \bibinfo {pages} {4953} (\bibinfo {year}
  {2016})}\BibitemShut {NoStop}%
\bibitem [{\citenamefont {Pierleoni}\ \emph {et~al.}(2018)\citenamefont
  {Pierleoni}, \citenamefont {Rillo}, \citenamefont {Ceperley},\ and\
  \citenamefont {Holzmann}}]{Pierleoni_2018}%
  \BibitemOpen
  \bibfield  {author} {\bibinfo {author} {\bibfnamefont {C.}~\bibnamefont
  {Pierleoni}}, \bibinfo {author} {\bibfnamefont {G.}~\bibnamefont {Rillo}},
  \bibinfo {author} {\bibfnamefont {D.~M.}\ \bibnamefont {Ceperley}}, \ and\
  \bibinfo {author} {\bibfnamefont {M.}~\bibnamefont {Holzmann}},\ }\href
  {\doibase 10.1088/1742-6596/1136/1/012005} {\bibfield  {journal} {\bibinfo
  {journal} {Journal of Physics: Conference Series}\ }\textbf {\bibinfo
  {volume} {1136}},\ \bibinfo {pages} {012005} (\bibinfo {year}
  {2018})}\BibitemShut {NoStop}%
\bibitem [{\citenamefont {Zaghoo}\ and\ \citenamefont
  {Silvera}(2017)}]{Zaghoo_2017}%
  \BibitemOpen
  \bibfield  {author} {\bibinfo {author} {\bibfnamefont {M.}~\bibnamefont
  {Zaghoo}}\ and\ \bibinfo {author} {\bibfnamefont {I.~F.}\ \bibnamefont
  {Silvera}},\ }\href {\doibase 10.1073/pnas.1707918114} {\bibfield  {journal}
  {\bibinfo  {journal} {Proceedings of the National Academy of Sciences}\
  }\textbf {\bibinfo {volume} {114}},\ \bibinfo {pages} {11873} (\bibinfo
  {year} {2017})}\BibitemShut {NoStop}%
\bibitem [{\citenamefont {Lee}\ and\ \citenamefont
  {Elliott}(2017)}]{lee2017the}%
  \BibitemOpen
  \bibfield  {author} {\bibinfo {author} {\bibfnamefont {T.~H.}\ \bibnamefont
  {Lee}}\ and\ \bibinfo {author} {\bibfnamefont {S.~R.}\ \bibnamefont
  {Elliott}},\ }\href@noop {} {\bibfield  {journal} {\bibinfo  {journal} {Adv.
  Mater.}\ }\textbf {\bibinfo {volume} {29}},\ \bibinfo {pages} {1700814}
  (\bibinfo {year} {2017})}\BibitemShut {NoStop}%
\bibitem [{\citenamefont {Loke}\ \emph {et~al.}(2012)\citenamefont {Loke},
  \citenamefont {Lee}, \citenamefont {Wang}, \citenamefont {Shi}, \citenamefont
  {Zhao}, \citenamefont {Yeo}, \citenamefont {Chong},\ and\ \citenamefont
  {Elliott}}]{Loke1566}%
  \BibitemOpen
  \bibfield  {author} {\bibinfo {author} {\bibfnamefont {D.}~\bibnamefont
  {Loke}}, \bibinfo {author} {\bibfnamefont {T.~H.}\ \bibnamefont {Lee}},
  \bibinfo {author} {\bibfnamefont {W.~J.}\ \bibnamefont {Wang}}, \bibinfo
  {author} {\bibfnamefont {L.~P.}\ \bibnamefont {Shi}}, \bibinfo {author}
  {\bibfnamefont {R.}~\bibnamefont {Zhao}}, \bibinfo {author} {\bibfnamefont
  {Y.~C.}\ \bibnamefont {Yeo}}, \bibinfo {author} {\bibfnamefont {T.~C.}\
  \bibnamefont {Chong}}, \ and\ \bibinfo {author} {\bibfnamefont {S.~R.}\
  \bibnamefont {Elliott}},\ }\href@noop {} {\bibfield  {journal} {\bibinfo
  {journal} {Science}\ }\textbf {\bibinfo {volume} {336}},\ \bibinfo {pages}
  {1566} (\bibinfo {year} {2012})}\BibitemShut {NoStop}%
\bibitem [{\citenamefont {Skelton}\ \emph {et~al.}(2015)\citenamefont
  {Skelton}, \citenamefont {Loke}, \citenamefont {Lee},\ and\ \citenamefont
  {Elliott}}]{Skelton_2015}%
  \BibitemOpen
  \bibfield  {author} {\bibinfo {author} {\bibfnamefont {J.~M.}\ \bibnamefont
  {Skelton}}, \bibinfo {author} {\bibfnamefont {D.}~\bibnamefont {Loke}},
  \bibinfo {author} {\bibfnamefont {T.}~\bibnamefont {Lee}}, \ and\ \bibinfo
  {author} {\bibfnamefont {S.~R.}\ \bibnamefont {Elliott}},\ }\href@noop {}
  {\bibfield  {journal} {\bibinfo  {journal} {ACS Appl. Mater. Interfaces}\
  }\textbf {\bibinfo {volume} {7}},\ \bibinfo {pages} {14223} (\bibinfo {year}
  {2015})}\BibitemShut {NoStop}%
\bibitem [{\citenamefont {Zavabeti}\ \emph {et~al.}(2017)\citenamefont
  {Zavabeti}, \citenamefont {Ou}, \citenamefont {Carey}, \citenamefont {Syed},
  \citenamefont {Orrell-Trigg}, \citenamefont {Mayes}, \citenamefont {Xu},
  \citenamefont {Kavehei}, \citenamefont {O{\textquoteright}Mullane},
  \citenamefont {Kaner}, \citenamefont {Kalantar-zadeh},\ and\ \citenamefont
  {Daeneke}}]{Zavabeti332}%
  \BibitemOpen
  \bibfield  {author} {\bibinfo {author} {\bibfnamefont {A.}~\bibnamefont
  {Zavabeti}}, \bibinfo {author} {\bibfnamefont {J.~Z.}\ \bibnamefont {Ou}},
  \bibinfo {author} {\bibfnamefont {B.~J.}\ \bibnamefont {Carey}}, \bibinfo
  {author} {\bibfnamefont {N.}~\bibnamefont {Syed}}, \bibinfo {author}
  {\bibfnamefont {R.}~\bibnamefont {Orrell-Trigg}}, \bibinfo {author}
  {\bibfnamefont {E.~L.~H.}\ \bibnamefont {Mayes}}, \bibinfo {author}
  {\bibfnamefont {C.}~\bibnamefont {Xu}}, \bibinfo {author} {\bibfnamefont
  {O.}~\bibnamefont {Kavehei}}, \bibinfo {author} {\bibfnamefont {A.~P.}\
  \bibnamefont {O{\textquoteright}Mullane}}, \bibinfo {author} {\bibfnamefont
  {R.~B.}\ \bibnamefont {Kaner}}, \bibinfo {author} {\bibfnamefont
  {K.}~\bibnamefont {Kalantar-zadeh}}, \ and\ \bibinfo {author} {\bibfnamefont
  {T.}~\bibnamefont {Daeneke}},\ }\href@noop {} {\bibfield  {journal} {\bibinfo
   {journal} {Science}\ }\textbf {\bibinfo {volume} {358}},\ \bibinfo {pages}
  {332} (\bibinfo {year} {2017})}\BibitemShut {NoStop}%
\bibitem [{\citenamefont {{P}aul Poirier}(1994)}]{Poirier:1994}%
  \BibitemOpen
  \bibfield  {author} {\bibinfo {author} {\bibfnamefont {J.}~\bibnamefont
  {{P}aul Poirier}},\ }\href@noop {} {\bibfield  {journal} {\bibinfo  {journal}
  {Phys. Earth Planet. Inter.}\ }\textbf {\bibinfo {volume} {85}},\ \bibinfo
  {pages} {319} (\bibinfo {year} {1994})}\BibitemShut {NoStop}%
\bibitem [{\citenamefont {Pozzo}\ \emph {et~al.}(2013)\citenamefont {Pozzo},
  \citenamefont {Davies}, \citenamefont {Gubbins},\ and\ \citenamefont
  {Alf{\`e}}}]{Pozzo_2013}%
  \BibitemOpen
  \bibfield  {author} {\bibinfo {author} {\bibfnamefont {M.}~\bibnamefont
  {Pozzo}}, \bibinfo {author} {\bibfnamefont {C.}~\bibnamefont {Davies}},
  \bibinfo {author} {\bibfnamefont {D.}~\bibnamefont {Gubbins}}, \ and\
  \bibinfo {author} {\bibfnamefont {D.}~\bibnamefont {Alf{\`e}}},\ }\href
  {\doibase 10.1103/physrevb.87.014110} {\bibfield  {journal} {\bibinfo
  {journal} {Physical Review B}\ }\textbf {\bibinfo {volume} {87}} (\bibinfo
  {year} {2013}),\ 10.1103/physrevb.87.014110}\BibitemShut {NoStop}%
\bibitem [{\citenamefont {Pozzo}\ \emph {et~al.}(2014)\citenamefont {Pozzo},
  \citenamefont {Davies}, \citenamefont {Gubbins},\ and\ \citenamefont
  {Alf{\`e}}}]{Pozzo:2014}%
  \BibitemOpen
  \bibfield  {author} {\bibinfo {author} {\bibfnamefont {M.}~\bibnamefont
  {Pozzo}}, \bibinfo {author} {\bibfnamefont {C.}~\bibnamefont {Davies}},
  \bibinfo {author} {\bibfnamefont {D.}~\bibnamefont {Gubbins}}, \ and\
  \bibinfo {author} {\bibfnamefont {D.}~\bibnamefont {Alf{\`e}}},\ }\href@noop
  {} {\bibfield  {journal} {\bibinfo  {journal} {Earth Planet. Sci. Lett.}\
  }\textbf {\bibinfo {volume} {393}},\ \bibinfo {pages} {159} (\bibinfo {year}
  {2014})}\BibitemShut {NoStop}%
\bibitem [{\citenamefont {Williams}(2018)}]{Williams:2018}%
  \BibitemOpen
  \bibfield  {author} {\bibinfo {author} {\bibfnamefont {Q.}~\bibnamefont
  {Williams}},\ }\href@noop {} {\bibfield  {journal} {\bibinfo  {journal}
  {Annu. Rev. Earth Planet. Sci.}\ }\textbf {\bibinfo {volume} {46}},\ \bibinfo
  {pages} {47} (\bibinfo {year} {2018})}\BibitemShut {NoStop}%
\bibitem [{\citenamefont {Mazzola}, \citenamefont {Helled},\ and\ \citenamefont
  {Sorella}(2018)}]{Mazzola_2018}%
  \BibitemOpen
  \bibfield  {author} {\bibinfo {author} {\bibfnamefont {G.}~\bibnamefont
  {Mazzola}}, \bibinfo {author} {\bibfnamefont {R.}~\bibnamefont {Helled}}, \
  and\ \bibinfo {author} {\bibfnamefont {S.}~\bibnamefont {Sorella}},\ }\href
  {\doibase 10.1103/physrevlett.120.025701} {\bibfield  {journal} {\bibinfo
  {journal} {Physical Review Letters}\ }\textbf {\bibinfo {volume} {120}}
  (\bibinfo {year} {2018}),\ 10.1103/physrevlett.120.025701}\BibitemShut
  {NoStop}%
\bibitem [{\citenamefont {Sugino}\ and\ \citenamefont
  {Car}(1995)}]{Sugino:PRL:1995}%
  \BibitemOpen
  \bibfield  {author} {\bibinfo {author} {\bibfnamefont {O.}~\bibnamefont
  {Sugino}}\ and\ \bibinfo {author} {\bibfnamefont {R.}~\bibnamefont {Car}},\
  }\href@noop {} {\bibfield  {journal} {\bibinfo  {journal} {Phys. Rev. Lett.}\
  }\textbf {\bibinfo {volume} {74}},\ \bibinfo {pages} {1823} (\bibinfo {year}
  {1995})}\BibitemShut {NoStop}%
\bibitem [{\citenamefont {Alf{\`e}}\ and\ \citenamefont
  {Gillan}(2003)}]{Alfe:PRB:2003}%
  \BibitemOpen
  \bibfield  {author} {\bibinfo {author} {\bibfnamefont {D.}~\bibnamefont
  {Alf{\`e}}}\ and\ \bibinfo {author} {\bibfnamefont {M.~J.}\ \bibnamefont
  {Gillan}},\ }\href@noop {} {\bibfield  {journal} {\bibinfo  {journal} {Phys.
  Rev. B}\ }\textbf {\bibinfo {volume} {68}},\ \bibinfo {pages} {205212}
  (\bibinfo {year} {2003})}\BibitemShut {NoStop}%
\bibitem [{\citenamefont {McMillan}\ \emph {et~al.}(2005)\citenamefont
  {McMillan}, \citenamefont {Wilson}, \citenamefont {Daisenberger},\ and\
  \citenamefont {Machon}}]{McMillan:2005ij}%
  \BibitemOpen
  \bibfield  {author} {\bibinfo {author} {\bibfnamefont {P.~F.}\ \bibnamefont
  {McMillan}}, \bibinfo {author} {\bibfnamefont {M.}~\bibnamefont {Wilson}},
  \bibinfo {author} {\bibfnamefont {D.}~\bibnamefont {Daisenberger}}, \ and\
  \bibinfo {author} {\bibfnamefont {D.}~\bibnamefont {Machon}},\ }\href
  {\doibase 10.1038/nmat1458} {\bibfield  {journal} {\bibinfo  {journal} {Nat
  Mater}\ }\textbf {\bibinfo {volume} {4}},\ \bibinfo {pages} {680} (\bibinfo
  {year} {2005})}\BibitemShut {NoStop}%
\bibitem [{\citenamefont {Ganesh}\ and\ \citenamefont
  {Widom}(2009)}]{Ganesh:2009om}%
  \BibitemOpen
  \bibfield  {author} {\bibinfo {author} {\bibfnamefont {P.}~\bibnamefont
  {Ganesh}}\ and\ \bibinfo {author} {\bibfnamefont {M.}~\bibnamefont {Widom}},\
  }\href {\doibase 10.1103/PhysRevLett.102.075701} {\bibfield  {journal}
  {\bibinfo  {journal} {Phys Rev Lett}\ }\textbf {\bibinfo {volume} {102}},\
  \bibinfo {pages} {075701} (\bibinfo {year} {2009})}\BibitemShut {NoStop}%
\bibitem [{\citenamefont {Beye}\ \emph {et~al.}(2010)\citenamefont {Beye},
  \citenamefont {Sorgenfrei}, \citenamefont {Schlotter}, \citenamefont
  {Wurth},\ and\ \citenamefont {F{\"o}hlisch}}]{Beye:2010oq}%
  \BibitemOpen
  \bibfield  {author} {\bibinfo {author} {\bibfnamefont {M.}~\bibnamefont
  {Beye}}, \bibinfo {author} {\bibfnamefont {F.}~\bibnamefont {Sorgenfrei}},
  \bibinfo {author} {\bibfnamefont {W.~F.}\ \bibnamefont {Schlotter}}, \bibinfo
  {author} {\bibfnamefont {W.}~\bibnamefont {Wurth}}, \ and\ \bibinfo {author}
  {\bibfnamefont {A.}~\bibnamefont {F{\"o}hlisch}},\ }\href {\doibase
  10.1073/pnas.1006499107} {\bibfield  {journal} {\bibinfo  {journal} {Proc
  Natl Acad Sci U S A}\ }\textbf {\bibinfo {volume} {107}},\ \bibinfo {pages}
  {16772} (\bibinfo {year} {2010})}\BibitemShut {NoStop}%
\bibitem [{\citenamefont {Sastry}(2010)}]{Sastry:2010kl}%
  \BibitemOpen
  \bibfield  {author} {\bibinfo {author} {\bibfnamefont {S.}~\bibnamefont
  {Sastry}},\ }\href {\doibase 10.1073/pnas.1012192107} {\bibfield  {journal}
  {\bibinfo  {journal} {Proc Natl Acad Sci U S A}\ }\textbf {\bibinfo {volume}
  {107}},\ \bibinfo {pages} {17063} (\bibinfo {year} {2010})}\BibitemShut
  {NoStop}%
\bibitem [{\citenamefont {VandeVondele}\ \emph {et~al.}(2005)\citenamefont
  {VandeVondele}, \citenamefont {Krack}, \citenamefont {Mohamed}, \citenamefont
  {Parrinello}, \citenamefont {Chassaing},\ and\ \citenamefont
  {Hutter}}]{VandeVondele2005}%
  \BibitemOpen
  \bibfield  {author} {\bibinfo {author} {\bibfnamefont {J.}~\bibnamefont
  {VandeVondele}}, \bibinfo {author} {\bibfnamefont {M.}~\bibnamefont {Krack}},
  \bibinfo {author} {\bibfnamefont {F.}~\bibnamefont {Mohamed}}, \bibinfo
  {author} {\bibfnamefont {M.}~\bibnamefont {Parrinello}}, \bibinfo {author}
  {\bibfnamefont {T.}~\bibnamefont {Chassaing}}, \ and\ \bibinfo {author}
  {\bibfnamefont {J.}~\bibnamefont {Hutter}},\ }\href {\doibase DOI
  10.1016/j.cpc.2004.12.014} {\bibfield  {journal} {\bibinfo  {journal}
  {Comput.\ Phys.\ Commun.}\ }\textbf {\bibinfo {volume} {167}},\ \bibinfo
  {pages} {103} (\bibinfo {year} {2005})}\BibitemShut {NoStop}%
\bibitem [{\citenamefont {VandeVondele}\ and\ \citenamefont
  {Hutter}(2007)}]{VandeVondele2007}%
  \BibitemOpen
  \bibfield  {author} {\bibinfo {author} {\bibfnamefont {J.}~\bibnamefont
  {VandeVondele}}\ and\ \bibinfo {author} {\bibfnamefont {J.}~\bibnamefont
  {Hutter}},\ }\href {\doibase DOI 10.1063/1.2770708} {\bibfield  {journal}
  {\bibinfo  {journal} {J.\ Chem.\ Phys.}\ }\textbf {\bibinfo {volume} {127}},\
  \bibinfo {pages} {114105} (\bibinfo {year} {2007})}\BibitemShut {NoStop}%
\bibitem [{\citenamefont {Goedecker}, \citenamefont {Teter},\ and\
  \citenamefont {Hutter}(1996)}]{Goedecker1996}%
  \BibitemOpen
  \bibfield  {author} {\bibinfo {author} {\bibfnamefont {S.}~\bibnamefont
  {Goedecker}}, \bibinfo {author} {\bibfnamefont {M.}~\bibnamefont {Teter}}, \
  and\ \bibinfo {author} {\bibfnamefont {J.}~\bibnamefont {Hutter}},\
  }\href@noop {} {\bibfield  {journal} {\bibinfo  {journal} {Phys.\ Rev.\ B}\
  }\textbf {\bibinfo {volume} {54}},\ \bibinfo {pages} {1703} (\bibinfo {year}
  {1996})}\BibitemShut {NoStop}%
\bibitem [{\citenamefont {Sun}, \citenamefont {Ruzsinszky},\ and\ \citenamefont
  {Perdew}(2015)}]{SCAN}%
  \BibitemOpen
  \bibfield  {author} {\bibinfo {author} {\bibfnamefont {J.}~\bibnamefont
  {Sun}}, \bibinfo {author} {\bibfnamefont {A.}~\bibnamefont {Ruzsinszky}}, \
  and\ \bibinfo {author} {\bibfnamefont {J.~P.}\ \bibnamefont {Perdew}},\
  }\href@noop {} {\bibfield  {journal} {\bibinfo  {journal} {Phys. Rev. Lett.}\
  }\textbf {\bibinfo {volume} {115}},\ \bibinfo {pages} {036402} (\bibinfo
  {year} {2015})}\BibitemShut {NoStop}%
\bibitem [{\citenamefont {Lehtola}\ \emph {et~al.}(2018)\citenamefont
  {Lehtola}, \citenamefont {Steigemann}, \citenamefont {Oliveira},\ and\
  \citenamefont {Marques}}]{LEHTOLA20181}%
  \BibitemOpen
  \bibfield  {author} {\bibinfo {author} {\bibfnamefont {S.}~\bibnamefont
  {Lehtola}}, \bibinfo {author} {\bibfnamefont {C.}~\bibnamefont {Steigemann}},
  \bibinfo {author} {\bibfnamefont {M.~J.}\ \bibnamefont {Oliveira}}, \ and\
  \bibinfo {author} {\bibfnamefont {M.~A.}\ \bibnamefont {Marques}},\ }\href
  {\doibase https://doi.org/10.1016/j.softx.2017.11.002} {\bibfield  {journal}
  {\bibinfo  {journal} {SoftwareX}\ }\textbf {\bibinfo {volume} {7}},\ \bibinfo
  {pages} {1 } (\bibinfo {year} {2018})}\BibitemShut {NoStop}%
\bibitem [{\citenamefont {Marques}, \citenamefont {Oliveira},\ and\
  \citenamefont {Burnus}(2012)}]{MARQUES20122272}%
  \BibitemOpen
  \bibfield  {author} {\bibinfo {author} {\bibfnamefont {M.~A.}\ \bibnamefont
  {Marques}}, \bibinfo {author} {\bibfnamefont {M.~J.}\ \bibnamefont
  {Oliveira}}, \ and\ \bibinfo {author} {\bibfnamefont {T.}~\bibnamefont
  {Burnus}},\ }\href {\doibase https://doi.org/10.1016/j.cpc.2012.05.007}
  {\bibfield  {journal} {\bibinfo  {journal} {Computer Physics Communications}\
  }\textbf {\bibinfo {volume} {183}},\ \bibinfo {pages} {2272 } (\bibinfo
  {year} {2012})}\BibitemShut {NoStop}%
\bibitem [{\citenamefont {Bussi}, \citenamefont {Donadio},\ and\ \citenamefont
  {Parrinello}(2007)}]{Bussi:JCP:2007}%
  \BibitemOpen
  \bibfield  {author} {\bibinfo {author} {\bibfnamefont {G.}~\bibnamefont
  {Bussi}}, \bibinfo {author} {\bibfnamefont {D.}~\bibnamefont {Donadio}}, \
  and\ \bibinfo {author} {\bibfnamefont {M.}~\bibnamefont {Parrinello}},\
  }\href@noop {} {\bibfield  {journal} {\bibinfo  {journal} {J. Chem. Phys.}\
  }\textbf {\bibinfo {volume} {126}},\ \bibinfo {pages} {014101} (\bibinfo
  {year} {2007})}\BibitemShut {NoStop}%
\bibitem [{\citenamefont {Marzari}\ \emph {et~al.}(2012)\citenamefont
  {Marzari}, \citenamefont {Mostofi}, \citenamefont {Yates}, \citenamefont
  {Souza},\ and\ \citenamefont {Vanderbilt}}]{RevModPhys.84.1419}%
  \BibitemOpen
  \bibfield  {author} {\bibinfo {author} {\bibfnamefont {N.}~\bibnamefont
  {Marzari}}, \bibinfo {author} {\bibfnamefont {A.~A.}\ \bibnamefont
  {Mostofi}}, \bibinfo {author} {\bibfnamefont {J.~R.}\ \bibnamefont {Yates}},
  \bibinfo {author} {\bibfnamefont {I.}~\bibnamefont {Souza}}, \ and\ \bibinfo
  {author} {\bibfnamefont {D.}~\bibnamefont {Vanderbilt}},\ }\href {\doibase
  10.1103/RevModPhys.84.1419} {\bibfield  {journal} {\bibinfo  {journal} {Rev.
  Mod. Phys.}\ }\textbf {\bibinfo {volume} {84}},\ \bibinfo {pages} {1419}
  (\bibinfo {year} {2012})}\BibitemShut {NoStop}%
\bibitem [{\citenamefont {Silvestrelli}\ \emph {et~al.}(1998)\citenamefont
  {Silvestrelli}, \citenamefont {Marzari}, \citenamefont {Vanderbilt},\ and\
  \citenamefont {Parrinello}}]{Silvestrelli_SSC_1998}%
  \BibitemOpen
  \bibfield  {author} {\bibinfo {author} {\bibfnamefont {P.~L.}\ \bibnamefont
  {Silvestrelli}}, \bibinfo {author} {\bibfnamefont {N.}~\bibnamefont
  {Marzari}}, \bibinfo {author} {\bibfnamefont {D.}~\bibnamefont {Vanderbilt}},
  \ and\ \bibinfo {author} {\bibfnamefont {M.}~\bibnamefont {Parrinello}},\
  }\href@noop {} {\bibfield  {journal} {\bibinfo  {journal} {Solid State
  Commun.}\ }\textbf {\bibinfo {volume} {107}},\ \bibinfo {pages} {7} (\bibinfo
  {year} {1998})}\BibitemShut {NoStop}%
\bibitem [{\citenamefont {Remsing}\ and\ \citenamefont
  {Klein}(2019)}]{Remsing:JPCB:2019}%
  \BibitemOpen
  \bibfield  {author} {\bibinfo {author} {\bibfnamefont {R.~C.}\ \bibnamefont
  {Remsing}}\ and\ \bibinfo {author} {\bibfnamefont {M.~L.}\ \bibnamefont
  {Klein}},\ }\href@noop {} {\bibfield  {journal} {\bibinfo  {journal} {J.
  Phys. Chem. B}\ }\textbf {\bibinfo {volume} {123}},\ \bibinfo {pages} {6266}
  (\bibinfo {year} {2019})}\BibitemShut {NoStop}%
\bibitem [{\citenamefont {Remsing}\ \emph {et~al.}(2018)\citenamefont
  {Remsing}, \citenamefont {Sun}, \citenamefont {Waghmare},\ and\ \citenamefont
  {Klein}}]{Remsing:MolPhys:2018}%
  \BibitemOpen
  \bibfield  {author} {\bibinfo {author} {\bibfnamefont {R.~C.}\ \bibnamefont
  {Remsing}}, \bibinfo {author} {\bibfnamefont {J.}~\bibnamefont {Sun}},
  \bibinfo {author} {\bibfnamefont {U.~V.}\ \bibnamefont {Waghmare}}, \ and\
  \bibinfo {author} {\bibfnamefont {M.~L.}\ \bibnamefont {Klein}},\ }\href@noop
  {} {\bibfield  {journal} {\bibinfo  {journal} {Mol. Phys.}\ }\textbf
  {\bibinfo {volume} {116}},\ \bibinfo {pages} {3372} (\bibinfo {year}
  {2018})}\BibitemShut {NoStop}%
\bibitem [{\citenamefont {Remsing}\ and\ \citenamefont
  {Klein}(2020)}]{Remsing:PRL:2020}%
  \BibitemOpen
  \bibfield  {author} {\bibinfo {author} {\bibfnamefont {R.~C.}\ \bibnamefont
  {Remsing}}\ and\ \bibinfo {author} {\bibfnamefont {M.~L.}\ \bibnamefont
  {Klein}},\ }\href@noop {} {\bibfield  {journal} {\bibinfo  {journal} {Phys.
  Rev. Lett.}\ }\textbf {\bibinfo {volume} {124}},\ \bibinfo {pages} {066001}
  (\bibinfo {year} {2020})}\BibitemShut {NoStop}%
\bibitem [{\citenamefont {Blount}(1962)}]{blount1962formalisms}%
  \BibitemOpen
  \bibfield  {author} {\bibinfo {author} {\bibfnamefont {E.}~\bibnamefont
  {Blount}},\ }in\ \href@noop {} {\emph {\bibinfo {booktitle} {Solid state
  physics}}},\ Vol.~\bibinfo {volume} {13}\ (\bibinfo  {publisher} {Elsevier},\
  \bibinfo {year} {1962})\ pp.\ \bibinfo {pages} {305--373}\BibitemShut
  {NoStop}%
\bibitem [{\citenamefont {Chandler}(1978)}]{Chandler_1978}%
  \BibitemOpen
  \bibfield  {author} {\bibinfo {author} {\bibfnamefont {D.}~\bibnamefont
  {Chandler}},\ }\href@noop {} {\bibfield  {journal} {\bibinfo  {journal} {J.
  Chem. Phys.}\ }\textbf {\bibinfo {volume} {68}},\ \bibinfo {pages} {2959}
  (\bibinfo {year} {1978})}\BibitemShut {NoStop}%
\bibitem [{\citenamefont {Luzar}(2000)}]{Luzar_2000}%
  \BibitemOpen
  \bibfield  {author} {\bibinfo {author} {\bibfnamefont {A.}~\bibnamefont
  {Luzar}},\ }\href@noop {} {\bibfield  {journal} {\bibinfo  {journal} {J.
  Chem. Phys.}\ }\textbf {\bibinfo {volume} {113}},\ \bibinfo {pages} {10663}
  (\bibinfo {year} {2000})}\BibitemShut {NoStop}%
\bibitem [{\citenamefont {Luzar}\ and\ \citenamefont
  {Chandler}(1996)}]{LuzarChandler}%
  \BibitemOpen
  \bibfield  {author} {\bibinfo {author} {\bibfnamefont {A.}~\bibnamefont
  {Luzar}}\ and\ \bibinfo {author} {\bibfnamefont {D.}~\bibnamefont
  {Chandler}},\ }\href@noop {} {\bibfield  {journal} {\bibinfo  {journal}
  {Nature}\ }\textbf {\bibinfo {volume} {379}},\ \bibinfo {pages} {55}
  (\bibinfo {year} {1996})}\BibitemShut {NoStop}%
\bibitem [{\citenamefont {Stiffler}, \citenamefont {Thompson},\ and\
  \citenamefont {Peercy}(1988)}]{Stiffler:PRL:1988}%
  \BibitemOpen
  \bibfield  {author} {\bibinfo {author} {\bibfnamefont {S.~R.}\ \bibnamefont
  {Stiffler}}, \bibinfo {author} {\bibfnamefont {M.~O.}\ \bibnamefont
  {Thompson}}, \ and\ \bibinfo {author} {\bibfnamefont {P.~S.}\ \bibnamefont
  {Peercy}},\ }\href@noop {} {\bibfield  {journal} {\bibinfo  {journal} {Phys
  Rev Lett}\ }\textbf {\bibinfo {volume} {60}},\ \bibinfo {pages} {2519}
  (\bibinfo {year} {1988})}\BibitemShut {NoStop}%
\bibitem [{\citenamefont {Sanders}\ and\ \citenamefont
  {Aziz}(1999)}]{Sanders:JApplPhys:1999}%
  \BibitemOpen
  \bibfield  {author} {\bibinfo {author} {\bibfnamefont {P.~G.}\ \bibnamefont
  {Sanders}}\ and\ \bibinfo {author} {\bibfnamefont {M.~J.}\ \bibnamefont
  {Aziz}},\ }\href@noop {} {\bibfield  {journal} {\bibinfo  {journal} {J. Appl.
  Phys.}\ }\textbf {\bibinfo {volume} {86}},\ \bibinfo {pages} {4258} (\bibinfo
  {year} {1999})}\BibitemShut {NoStop}%
\bibitem [{\citenamefont {Wagner}, \citenamefont {Su},\ and\ \citenamefont
  {Grobe}(2010)}]{Wagner_2010}%
  \BibitemOpen
  \bibfield  {author} {\bibinfo {author} {\bibfnamefont {R.~E.}\ \bibnamefont
  {Wagner}}, \bibinfo {author} {\bibfnamefont {Q.}~\bibnamefont {Su}}, \ and\
  \bibinfo {author} {\bibfnamefont {R.}~\bibnamefont {Grobe}},\ }\href@noop {}
  {\bibfield  {journal} {\bibinfo  {journal} {Phys. Rev. A}\ }\textbf {\bibinfo
  {volume} {82}},\ \bibinfo {pages} {022719} (\bibinfo {year}
  {2010})}\BibitemShut {NoStop}%
\bibitem [{\citenamefont {Kemper}\ \emph {et~al.}(2013)\citenamefont {Kemper},
  \citenamefont {Sentef}, \citenamefont {Moritz}, \citenamefont {Kao},
  \citenamefont {Shen}, \citenamefont {Freericks},\ and\ \citenamefont
  {Devereaux}}]{Kemper_2013}%
  \BibitemOpen
  \bibfield  {author} {\bibinfo {author} {\bibfnamefont {A.~F.}\ \bibnamefont
  {Kemper}}, \bibinfo {author} {\bibfnamefont {M.}~\bibnamefont {Sentef}},
  \bibinfo {author} {\bibfnamefont {B.}~\bibnamefont {Moritz}}, \bibinfo
  {author} {\bibfnamefont {C.~C.}\ \bibnamefont {Kao}}, \bibinfo {author}
  {\bibfnamefont {Z.~X.}\ \bibnamefont {Shen}}, \bibinfo {author}
  {\bibfnamefont {J.~K.}\ \bibnamefont {Freericks}}, \ and\ \bibinfo {author}
  {\bibfnamefont {T.~P.}\ \bibnamefont {Devereaux}},\ }\href@noop {} {\bibfield
   {journal} {\bibinfo  {journal} {Phys. Rev. B}\ }\textbf {\bibinfo {volume}
  {87}},\ \bibinfo {pages} {235139} (\bibinfo {year} {2013})}\BibitemShut
  {NoStop}%
\bibitem [{\citenamefont {Grosser}, \citenamefont {Slowik},\ and\ \citenamefont
  {Santra}(2017)}]{Grosser_2017}%
  \BibitemOpen
  \bibfield  {author} {\bibinfo {author} {\bibfnamefont {M.}~\bibnamefont
  {Grosser}}, \bibinfo {author} {\bibfnamefont {J.~M.}\ \bibnamefont {Slowik}},
  \ and\ \bibinfo {author} {\bibfnamefont {R.}~\bibnamefont {Santra}},\
  }\href@noop {} {\bibfield  {journal} {\bibinfo  {journal} {Phys. Rev. A}\
  }\textbf {\bibinfo {volume} {95}},\ \bibinfo {pages} {062107} (\bibinfo
  {year} {2017})}\BibitemShut {NoStop}%
\bibitem [{\citenamefont {Bryant}, \citenamefont {Johnson},\ and\ \citenamefont
  {Rossky}(2012)}]{bryant2012water}%
  \BibitemOpen
  \bibfield  {author} {\bibinfo {author} {\bibfnamefont {R.~G.}\ \bibnamefont
  {Bryant}}, \bibinfo {author} {\bibfnamefont {M.~A.}\ \bibnamefont {Johnson}},
  \ and\ \bibinfo {author} {\bibfnamefont {P.~J.}\ \bibnamefont {Rossky}},\
  }\href@noop {} {\bibfield  {journal} {\bibinfo  {journal} {Acc. Chem. Res.}\
  }\textbf {\bibinfo {volume} {45}},\ \bibinfo {pages} {1} (\bibinfo {year}
  {2012})}\BibitemShut {NoStop}%
\end{thebibliography}%

\end{document}